\documentclass[twocolumn,
                showpacs,
                nofootinbib,
                nobibnotes,
                aps,
                superscriptaddress,
                amssymb,
                amsmath,
                floatfix,
                reprint,
                prb,
                notitlepage,
                showkeys,
                10pt,
		]{revtex4-2}

\usepackage{amssymb}
\usepackage{amsmath}
\usepackage{cprotect}
\usepackage{graphicx}
\usepackage{dcolumn}
\usepackage{url}
\usepackage[colorlinks=true,breaklinks=true,allcolors=blue]{hyperref}
\usepackage{tikz}
\usepackage{dsfont}
\usepackage{epsfig}
\usepackage{color}
\usepackage[ruled]{algorithm2e}
\usepackage{bbold}
\usepackage[binary-units=true]{siunitx}
\usepackage{glossaries}
\usepackage{tikz}
\usepackage{bm}
\usepackage{hhline}
\usepackage{spverbatim}
\usepackage{mhchem}
\usepackage{siunitx}

\makeatletter
\let\cat@comma@active\@empty
\makeatother

\newcommand{\Eq}[1]{Eq.\,\eqref{eq:#1}}
\newcommand{\Fig}[1]{Fig.~\ref{fig:#1}}

\newcommand{\ket}[1]{|#1\rangle}
\newcommand{\bra}[1]{\langle#1|}

\begin{document}
\title{Data-Enhanced Variational Monte Carlo Simulations for Rydberg Atom Arrays}

\author{Stefanie Czischek}
\email{sczischek@uwaterloo.ca}
\affiliation{Department of Physics and Astronomy, University of Waterloo, Ontario, N2L 3G1, Canada}

\author{M. Schuyler Moss}
\affiliation{Department of Physics and Astronomy, University of Waterloo, Ontario, N2L 3G1, Canada}

\author{Matthew Radzihovsky}
\thanks{Currently at Apple Inc. One Apple Park Way, Cupertino, CA 95014, USA}
\affiliation{Department of Physics, Stanford University, Stanford, CA 93405, USA}

\author{Ejaaz Merali}
\affiliation{Department of Physics and Astronomy, University of Waterloo, Ontario, N2L 3G1, Canada}
\affiliation{Perimeter Institute for Theoretical Physics, Waterloo, Ontario, N2L 2Y5, Canada}

\author{Roger G. Melko}
\affiliation{Department of Physics and Astronomy, University of Waterloo, Ontario, N2L 3G1, Canada}
\affiliation{Perimeter Institute for Theoretical Physics, Waterloo, Ontario, N2L 2Y5, Canada}

\date{\today}

\begin{abstract}
Rydberg atom arrays are programmable quantum simulators capable of preparing interacting qubit systems in a variety of quantum states. Due to long 
experimental preparation times, obtaining projective measurement data can be relatively slow for large arrays, which poses a challenge for state reconstruction methods 
such as tomography.
Today, novel groundstate wavefunction ans\"{a}tze like recurrent neural networks (RNNs) can be efficiently trained not only from projective measurement data, 
but also through Hamiltonian-guided variational Monte Carlo (VMC). 
In this paper, we demonstrate how pretraining modern RNNs on even small amounts of data significantly reduces the convergence time for a subsequent variational optimization
of the wavefunction.  
This suggests that essentially any amount of measurements obtained from a state prepared in an experimental quantum simulator 
could provide significant value for neural-network-based VMC strategies.
\end{abstract}
\maketitle
%
%
\section{Introduction} 
Rydberg atom arrays are powerful candidates for high-quality
quantum simulation and computing platforms~\cite{Scholl2021,Cong2021,Levine2019}.
State-of-the-art experiments use optical tweezers to arrange and individually address atoms on arbitrary lattices~\cite{Endres2016, Barredo2018, Ebadi2021},
allowing them to strongly interact with a many-body Hamiltonian \cite{Fendley2004}.
The combination of complex lattice structures together with the precise tuning of inter-atomic interactions
has enabled the preparation of various novel phases and phase transitions~\cite{Ebadi2021, Semeghini2021, Miles2021},
whose continuing experimental exploration is supported by a suite of rapidly advancing numerical simulation
technologies~\cite{Samajdar2020, Samajdar2021, Kalinowski2021, Verresen2021,Merali2021}.

The quantum state of an array is probed with fluorescent imaging techniques,
which provide projective measurements in the Rydberg occupation basis~\cite{Endres2016, Ebadi2021, Xu2021}.
Since each measurement is destructive, the repetition rate at which they can be performed is limited by a number of factors,
in particular the preparation time of the target quantum state. The probabilistic loading of the array
requires a non-trivial rearrangement of atoms, resulting in repetition rates on the order of a few measurements per
second~\cite{Endres2016, Ebadi2021}, with exact time scales depending on the specific experimental setup.
Thus, data acquisition is limited, especially when compared to competing quantum simulation platforms,
such as ion traps which allow for hundreds of measurements per second~\cite{Monroe2021},
or superconducting circuits where orders of magnitude more measurements per second can be achieved \cite{Walter2017}.

Data acquisition rates have obvious consequences for state reconstruction and characterization,
as well as the direct estimation of operator expectation values, which suffer variances that scale inversely proportional
to the number of independent measurements. Recently, neural network wavefunctions
have been explored as tools for leveraging limited measurement data, as they provide a powerful ansatz
for representing quantum states with systematically tunable
expressivity~\cite{Torlai2016, Carleo2017, Torlai2018, Torlai2018a, Carrasquilla2019, Hibat-Allah2020, Ma2021}.
For example, standard generative models adopted from the field of machine learning have been used to
tomographically reconstruct quantum states~\cite{Torlai2020, Torlai2019, Torlai2020a, Neugebauer2020, Vogel1989, Cramer2010, Lanyon2017}
and have demonstrated the ability to significantly reduce the amount of
measurements required for the accurate reconstruction of operator expectation values~\cite{Torlai2020}.
In addition to their designed ability for {\em data-driven} learning, these ans\"{a}tze have the ability to
find ground state wavefunctions of a given Hamiltonian by variational energy minimization, via
the same {\em Hamiltonian-driven} training methods common in variational Monte Carlo (VMC)~\cite{Melko2019, Carrasquilla2021, Becca2017}.
Modern neural network strategies provide VMC ans\"{a}tze that can systematically be made powerful enough that
their expressiveness is no longer the limiting bottleneck.
Instead, the optimization process often requires long convergence times~\cite{Bukov2021, Westerhout2020, Valenti2022, Hibat-Allah2021, Morawetz2021}
and physics-inspired modifications of the network structure are sometimes needed to reach accuracies comparable to traditional
VMC approaches~\cite{Ferrari2019,Viteritti}.
In addition, the power of wisely chosen initializations to improve convergence has already been demonstrated in
traditional VMC methods~\cite{Pilati2019, Umrigar2007}.

In this work, we leverage these unique features of neural network wavefunctions to
explore the effect of combined data- and Hamiltonian-driven learning \cite{Bennewitz2021}.
Beginning with a randomly initialized recurrent neural network (RNN)~\cite{Hibat-Allah2020},
we first optimize network parameters using a limited amount of simulated~\cite{Sandvik2003, Merali2021, Kalinowski2021}
Rydberg occupation data drawn from a two-dimensional array in the vicinity of a quantum phase transition.
Then, we continue optimizing the network variationally, in the spirit of the recent work
by Carrasquilla and Torlai~\cite{Carrasquilla2021}.
We find a significant enhancement in variationally obtaining the ground state wavefunction by
pretraining the RNN on a limited amount of quantum data.

%
%
%
\begin{figure}
    \includegraphics[width=0.95\linewidth]{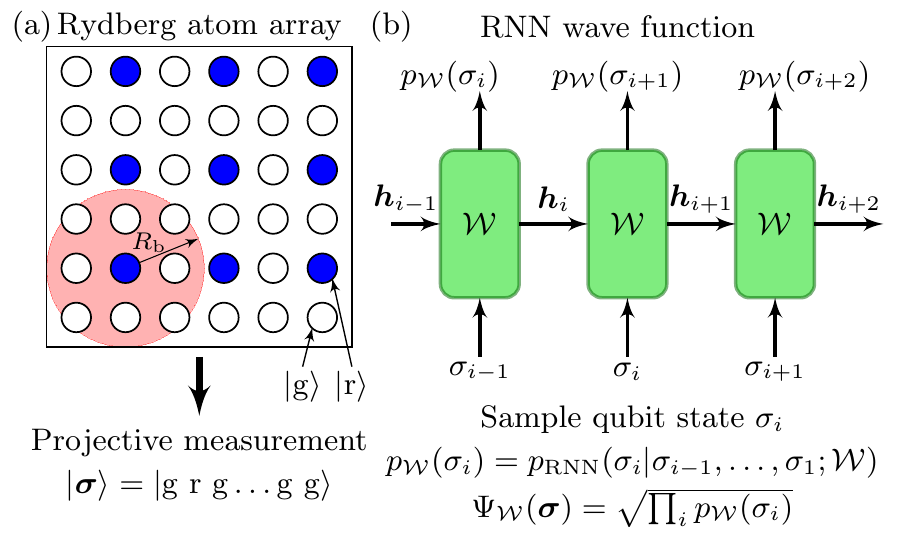}
    \caption{
        (a) Square lattice of $6\times 6$ atoms in the {\it striated} phase, where atoms in the Rydberg state (blue) are separated by
        stripes of atoms in the ground state (white) due to the Rydberg blockade radius $R_{\mathrm{b}}$.
        Projective measurements of individual atoms in the occupation number basis provide information about the
        quantum state.
        (b) A recurrent neural network (RNN), with green boxes representing a cell with
        $N_{\mathrm{h}}$ hidden units and tunable parameters $\mathcal{W}$.
        An input sequence $\boldsymbol{\sigma}$ is iteratively provided to the network, generating an output
        depending on the current input $\sigma_i$ and the hidden unit state $\boldsymbol{h}_i$.
        A single qubit state is used as input at iteration $i$
        and the network output is the probability distribution underlying the state of qubit $i+1$, from which
        the input $\sigma_{i+1}$ is sampled~\cite{Hibat-Allah2020}.
    }
    \label{fig:1}
\end{figure}
\section{Background}
We consider a system of $N=L\times L$ atoms arranged on a square lattice with
spacing $a=1$ and open boundary conditions [\Fig{1}(a)].
Each atom can be found in a ground state $\ket{\mathrm{g}}$ or in an excited (Rydberg) state $\ket{\mathrm{r}}$.
The system is driven by the many-body Hamiltonian,
\begin{align}
    \hat{H}&= -\frac{\Omega}{2}\sum_{i=1}^N\hat{\sigma}_i^{x}
    -\delta\sum_{i=1}^N\hat{n}_i+\sum_{i,j}V_{ij}\hat{n}_i\hat{n}_j,
    \label{eq:Hamiltonian}
\end{align}
with off-diagonal operator $\hat{\sigma}_i^x=\ket{\mathrm{g}}_i\bra{\mathrm{r}}_i$, and occupation operator
$\hat{n}_i=\ket{\mathrm{r}}_i\bra{\mathrm{r}}_i$ acting on atom $i$.
Atoms at positions $\boldsymbol{r}_i$ and $\boldsymbol{r}_j$ interact via the van der Waals potential
$V_{ij} = \Omega  R_{\mathrm{b}}^6 / {\left| \boldsymbol{r}_{i} - \boldsymbol{r}_{j} \right|}^6$,
where the Rydberg blockade radius $R_{\mathrm{b}}$ defines a region within which simultaneous excitations are
penalized~\cite{Jaksch2000, Lukin2001}.
Individual atoms are driven by a coherent laser with detuning $\delta$ and Rabi
frequency $\Omega$.
We fix $\delta=\Omega=1$ and $R_{\mathrm{b}}=7^{1/6}$,
so that $V_{ij}=7/|\boldsymbol{r}_{i}-\boldsymbol{r}_j|^6$, in the following.
This brings the system in the vicinity of the transition between the disordered and the striated phase
\cite{Ebadi2021}.

Information about the quantum state of a Rydberg array can be obtained from projective measurements in the occupation
basis, which gives full information about the positive real-valued ground state wavefunction of \Eq{Hamiltonian}~\cite{Endres2016, Ebadi2021, Xu2021}.
Each measurement forces a wavefunction collapse and the target state needs to be re-prepared before 
the next measurement can be performed.

Given sufficient projective measurements, a quantum state can be tomographically reconstructed,
e.g.~via a neural network wavefunction
ansatz~\cite{Torlai2020,Torlai2019,Torlai2018,Torlai2020a,Carrasquilla2021,Torlai2018a,Carrasquilla2019,
Ahmed2021,Cha2021,Neugebauer2020}.
Here, we focus on RNNs to represent quantum states, as 
illustrated in \Fig{1}(b), and choose the gated recurrent unit (GRU)~\cite{Cho2014} as network cell,
inspired by Refs.~\cite{Hibat-Allah2020, Carrasquilla2021}.
The RNN generates an output based on a sequence of inputs $\boldsymbol{\sigma}$ and the state of $N_{\mathrm{h}}$
hidden neurons per network cell.
During the training process, the network parameters $\mathcal{W}$ are tuned to generate a target 
output.
The amount of tunable parameters is defined by $N_{\mathrm{h}}$, which can be increased
to improve the network expressivity.
More details on RNNs are given in~\cite{Hibat-Allah2020, Carrasquilla2021, Cho2014}.

RNNs are naturally designed to encode probability distributions~\cite{Hibat-Allah2020}.
The network output at iteration $i$ can be interpreted as the conditional probability
distribution $p_{\mathrm{RNN}}(\sigma_{i}|\sigma_{i-1},\dots,\sigma_1;\mathcal{W})$, 
providing the joint distribution
$p_{\mathrm{RNN}}(\boldsymbol{\sigma};\mathcal{W}) =   \prod_i p_{\mathrm{RNN}}(\sigma_i|\sigma_{i-1},\dots,\sigma_1;\mathcal{W})$.
To represent a wavefunction, single qubit states are chosen as network input, which is iterated over the entire qubit
system.
The RNN is then trained to encode the probability distribution underlying projective measurements in the computational
basis.
In the case of positive real-valued wavefunctions, such as the ground state of \Eq{Hamiltonian},
the RNN represents the full quantum state
$\Psi_{\mathcal{W}}(\boldsymbol{\sigma}) =\langle\boldsymbol{\sigma}|\Psi\rangle= \sqrt{p_{\mathrm{RNN}}(\boldsymbol{\sigma};\mathcal{W})}$.
Further modifications can be used to reconstruct arbitrary complex
wavefunctions or density matrices~\cite{Hibat-Allah2020,Carrasquilla2021}.

Finally, samples from the generative step of the RNN-encoded distribution emulate projective measurement outcomes.
The quantum state of the full system can be sampled by iteratively drawing single qubit states.
We use these samples to evaluate the energy expectation value $H_{\mathrm{RNN}} = \left\langle\Psi_{\mathcal{W}}\right|\hat{H}\left|\Psi_{\mathcal{W}}\right\rangle$ via,
\begin{equation}
    H_{\mathrm{RNN}}
      =\sum_{\left\{\boldsymbol{\sigma}\right\}}\left|\Psi_{\mathcal{W}}\left(\boldsymbol{\sigma}\right)\right|^2
    H_{\mathrm{loc}}\left(\boldsymbol{\sigma}\right)
     \approx\frac{1}{N_{\mathrm{s}}}\sum_{\substack{\boldsymbol{\sigma}\sim\\
    p_{\mathrm{RNN}}\left(\boldsymbol{\sigma};\mathcal{W}\right)}}H_{\mathrm{loc}}\left(\boldsymbol{\sigma}\right),
\end{equation}
where we introduce the local energy,
\begin{align}
 H_{\mathrm{loc}}\left(\boldsymbol{\sigma}\right) &= \frac{ \left\langle\boldsymbol{\sigma}\right|\hat{H}
    \left|\Psi_{\mathcal{W}}\right\rangle }{ \left\langle\boldsymbol{\sigma}\right.\left|\Psi_{\mathcal{W}}\right\rangle},
    \label{eq:loc_Energy}
\end{align}
which is calculated and averaged over $N_{\mathrm{s}}$ samples $\boldsymbol{\sigma}$ drawn from
$p_{\mathrm{RNN}}(\boldsymbol{\sigma};\mathcal{W})$ and can efficiently be evaluated for local
non-diagonal operators~\cite{Carleo2017,Carrasquilla2021,Torlai2018,Melko2019,Hibat-Allah2020}.
%

\section{RNN training procedures}
We first explore the reconstruction of the ground state of a Rydberg atom array based on a
projective measurement data set $\mathcal{D}$.
In this data-driven setting, we optimize an RNN to approximate the
probability distribution $p_{\mathrm{RNN}}(\boldsymbol{\sigma})\approx p_{\mathcal{D}}(\boldsymbol{\sigma})$
underlying the data points $\boldsymbol{\sigma}\in\mathcal{D}$.
We use the Kullback-Leibler divergence
to define the loss function,
\begin{align}
    \mathcal{L}_{\mathrm{DKL}}\left(\mathcal{W}\right) &=\sum_{\left\{\boldsymbol{\sigma}\right\}}p_{\mathcal{D}}\left(\boldsymbol{\sigma}\right)\mathrm{log}\frac{p_{\mathcal{D}}
    \left(\boldsymbol{\sigma}\right)}{p_{\mathrm{RNN}}\left(\boldsymbol{\sigma};\mathcal{W}\right)}\\
    &\approx -S_{\mathcal{D}}-\frac{1}{\left|\mathcal{D}\right|}\sum_{\boldsymbol{\sigma}\in\mathcal{D}}
    \log p_{\mathrm{RNN}}\left(\boldsymbol{\sigma};\mathcal{W}\right).
    \label{eq:DKL}
\end{align}
In the last line we approximate $p_{\mathcal{D}}$ with the sum over the data in $\mathcal{D}$ and introduce the entropy
$S_{\mathcal{D}}=-\sum_{\left\{\boldsymbol{\sigma}\right\}}p_{\mathcal{D}}(\boldsymbol{\sigma})\mathrm{log}p_{\mathcal{D}}(\boldsymbol{\sigma})$.
We use the Adam optimizer~\cite{Kingma2014} to train a Glorot uniform initialized~\cite{Glorot2010} RNN by determining parameters $\mathcal{W}$ that minimize
the loss function.

\begin{figure}
    \includegraphics[width=\linewidth]{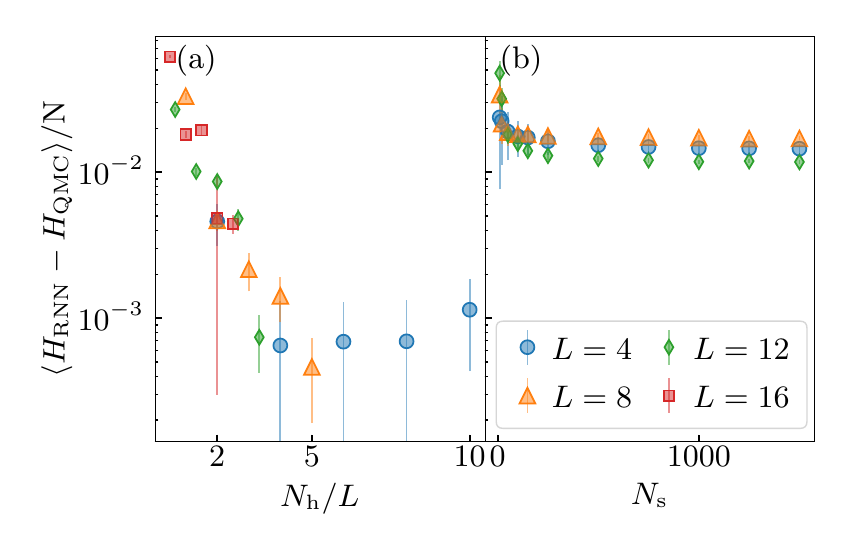}
    \caption{
        (a) Energy density difference of data-driven trained RNN states for different
        system sizes $N=L\times L$ as a function of the number of hidden neurons $N_{\mathrm{h}}$ per network cell.
        Results are averaged over iterations $1350$ to $1400$ ($950$ to $1000$ for $N=16\times 16$) and error bars denote standard deviations.
        The RNN is trained on $10^5$ data points with learning rate $\eta=10^{-4}$.
        (b) Energy density difference of an RNN trained with the Hamiltonian-driven procedure using $\eta=10^{-3}$.
        Results are plotted versus the number of RNN samples $N_{\mathrm{s}}$ for different system sizes, with $N_{\mathrm{h}}=2L$.
        Data is averaged over iterations $9500$ to $10{\small,}000$ with standard deviations as error bars.
        The QMC energy densities are $L=4$: $H_{\mathrm{QMC}}/16=-0.4534(1)$, $L=8$: $H_{\mathrm{QMC}}/64=-0.4052(2)$,
        $L=12$: $H_{\mathrm{QMC}}/144=-0.3885(2)$, $L=16$: $H_{\mathrm{QMC}}/256=-0.3805(2)$.
    }
    \label{fig:2}
\end{figure}

In order to produce a dataset $\mathcal{D}$, we use the quantum Monte Carlo (QMC) algorithm introduced in~\cite{Merali2021}, which is able to accurately emulate projective measurements of Rydberg atoms
in the ground state of the Hamiltonian in \Eq{Hamiltonian}.
We consider systems with up to $N=16\times 16$ atoms, relevant for state-of-the-art experimental
realizations~\cite{Ebadi2021}, for which an unbiased estimate of the energy $H_{\mathrm{QMC}}$ is easily obtained.
We generate $|\mathcal{D}|=10^5$ QMC samples for all considered system sizes, which gives estimates of the
ground state energy density $H_{\mathrm{QMC}}/N$ with errors on the order of $\sim 10^{-4}$ (caption of \Fig{2}).

We implement our RNN based on the code provided in~\cite{Carrasquilla2021, GitCode} and choose similar network
hyperparameters, fixing the learning rate to $\eta=0.001$ unless otherwise stated.
We first analyze the expressivity of the RNN in the data-driven approach to training.
In \Fig{2}(a) we train the network on the entire set of $|\mathcal{D}|=10^5$ QMC data points for different
system sizes of $N=L\times L$ atoms and plot the energy density difference
$\langle H_{\mathrm{RNN}}-H_{\mathrm{QMC}}\rangle/N$ as a function of the number of hidden neurons $N_{\mathrm{h}}$ per network cell.
The energy density differences are averaged over optimization iterations $1350$ to $1400$, where the training is approximately converged.
Each iteration corresponds to one training epoch where the full input dataset is given to the RNN in batches of $100$ randomly chosen samples and the network
parameters are updated.
The energy expectation value $H_{\mathrm{RNN}}$ is calculated on $1000$ samples drawn from
$p_{\mathrm{RNN}}(\boldsymbol{\sigma};\mathcal{W})$.
For $N=16\times 16$ we average the energy densities over iterations $950$ to $1000$ due to long computational runtimes, leading to larger variances
as convergence is not yet reached.
For all system sizes the energy density error shows a clear decreasing trend with increasing $N_{\mathrm{h}}$, which corresponds to higher network expressivity.
We find that the observed differences reach values below $10^{-2}$ for $N_{\mathrm{h}}/L\geq 2$ in all cases.
While higher reconstruction accuracies can be reached, increasing $N_{\mathrm{h}}$ comes at the price of higher
computational costs. We thus focus on $N_{\mathrm{h}}=2L$ as a practical compromise in the following.

Next, we train the RNN to represent the ground state of the same $N=L\times L$ system
using the Hamiltonian-driven approach.
In this procedure, the RNN parameters are optimized such that  
the energy expectation value $H_{\mathrm{RNN}}$ is minimized, corresponding to VMC.
With this motivation we define the loss function,
\begin{align}
    \mathcal{L}_{H}\left(\mathcal{W}\right)&=\frac{1}{N_{\mathrm{s}}}\sum_{\boldsymbol{\sigma}\sim
    p_{\mathrm{RNN}}\left(\boldsymbol{\sigma};\mathcal{W}\right)}H_{\mathrm{loc}}\left(\boldsymbol{\sigma}\right),
    \label{eq:loss}
\end{align}
and use again the Adam optimizer~\cite{Kingma2014} to find optimal network parameters $\mathcal{W}$.
Here we evaluate the local energy, \Eq{loc_Energy}, on
$N_{\mathrm{s}}$ samples drawn from $p_{\mathrm{RNN}}(\boldsymbol{\sigma};\mathcal{W})$.

In \Fig{2}(b) we consider the energy density difference as a
function of $N_{\mathrm{s}}$ for different system sizes.
The largest system size, $N=16\times 16$, is not shown due to exceeding computational runtimes.
We choose $N_{\mathrm{h}}=2L$ and average the measured energy densities over optimization iterations $9500$ to $10{\small,}000$.
The measured differences decrease with increasing $N_{\mathrm{s}}$, before they saturate at $\sim 10^{-2}$ for
$N_{\mathrm{s}}\geq 500$.
In accordance with~\cite{Carrasquilla2021} we thus fix $N_{\mathrm{s}}=1000$ in the following.
All system sizes show larger energy density differences than in \Fig{2}(a).
This demonstrates the limitation of the Hamiltonian-driven training procedure even after a large number of iterations,
which require prohibitively long computation times.
This large amount of required optimization steps is commonly caused by local optima in the parameter landscape and has led to
various model-inspired modifications of VMC~\cite{Umrigar2007, Pilati2019, Morawetz2021, Bukov2021, Valenti2022, Hibat-Allah2021}.
Below, we show that variational optimization can incur significant performance improvements without system-specific modifications,
by finding suitable initializations resulting from data-driven pretraining.

%
%
\begin{figure}
    \includegraphics[width=\linewidth]{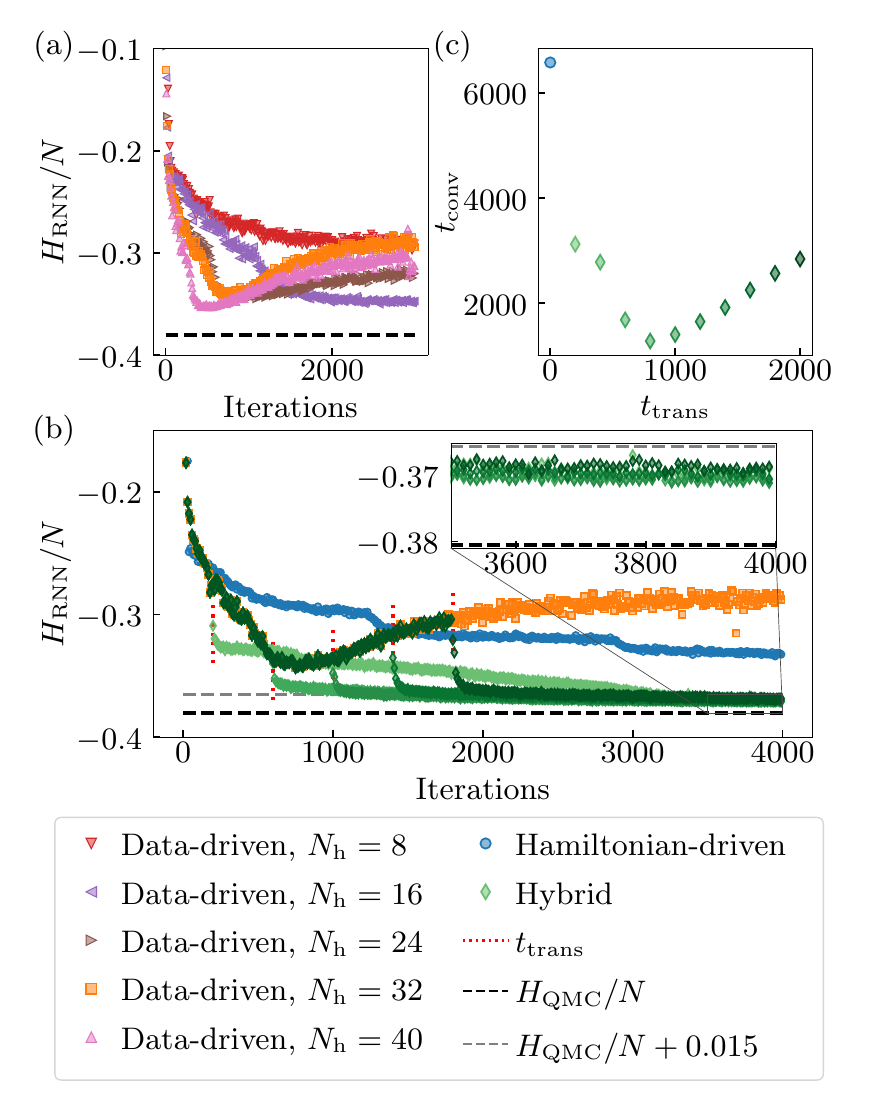}
    \caption{
        Quantum state reconstruction of an $N=16\times 16$ atom array.
        (a) Energy density $H_{\mathrm{RNN}}/N$ as a function of optimization iterations using the data-driven training
        on $1000$ data points.
        Different shapes and colors show different $N_{\mathrm{h}}$ compared to  
        the target energy density $H_{\mathrm{QMC}}/N$.
        (b) Energy density evolution during data-driven, Hamiltonian-driven, and hybrid
        training with different $t_{\mathrm{trans}}$ (darker green shows larger $t_{\mathrm{trans}}$) and $N_{\mathrm{h}}=32$.
        Inset emphasizes the convergence to the target solution after $\geq 3500$ iterations within a margin of $0.015$
        for all
        $t_{\mathrm{trans}}>0$.
        (c) Convergence time $t_{\mathrm{conv}}$ (see main text for definition)
        as a function of transition point
        $t_{\mathrm{trans}}$ in the hybrid training of an RNN with $N_{\mathrm{h}}=32$.
    }
    \label{fig:3}
\end{figure}
\section{Data-enhanced VMC}
In the above, we have shown that RNN wavefunctions can be made sufficiently expressive to represent the ground state of
large Rydberg arrays, however naive variational optimization of the neural network parameters
does not lead to accurate energies in a reasonable amount of simulation time.
Further, due to the long state-preparation times in modern Rydberg experiments \cite{Endres2016, Ebadi2021}, it is reasonable
to hypothesize that accurate data-driven reconstruction will be challenging on large arrays due to limited data.
To test this, we generate a randomly chosen subset of $\mathcal{D}$ containing $1000$ projective Rydberg occupation measurements, representing a
typical amount of experimental data.
In \Fig{3}(a) we show the energy density expectation value when training networks with different numbers of hidden
neurons $N_{\mathrm{h}}$ on the Rydberg ground state of $N=16\times 16$ atoms.
Instead of converging towards the estimated value, the energies for $N_{\mathrm{h}}\geq 24$ reach a minimum at $\sim 700$ iterations.
This phenomenon is easily recognized as overfitting.

Clearly, such limited datasets are insufficient for accurate state reconstruction. However, we now demonstrate the ability of small datasets
to enhance the performance of VMC in a \textit{hybrid} training procedure, defined with a simple change of loss function.
Namely, we begin training the RNN with the data-driven loss function \Eq{DKL}, before switching to Hamiltonian-driven training via \Eq{loss} after $t_{\mathrm{trans}}$ iterations.
\Fig{3}(b) illustrates the effectiveness of this simple hybrid procedure.
The green diamonds show the evolution of the energy using $1000$ data points in data-driven training,
before switching to Hamiltonian-driven training after $t_{\mathrm{trans}}$ iterations.
As a comparison, we also plot purely Hamiltonian-driven training results (i.e.~$t_{\mathrm{trans}}=0$).
We explore a number of different choices of $t_{\mathrm{trans}}$, which show significantly better
convergence than the pure Hamiltonian-driven variational method out to $4000$ optimization steps. 
All hybrid-trained simulations reach similar energy densities at $\gtrsim 3500$ iterations, 
which approximate the estimated value within a margin of $0.015$ (see \Fig{3}(b) inset).

In order to quantify the performance improvement of our algorithm,
we define a convergence time $t_{\mathrm{conv}}$ as the iteration after which the energy density difference reaches
$(H_{\mathrm{RNN}}-H_{\mathrm{QMC}})/N\leq 0.015$ for the first time.
Results are illustrated in \Fig{3}(c), where for better accuracy we consider the running average over $50$ iterations,
$\frac{1}{50}\sum_{i=-24}^{25}H_{\mathrm{RNN}}(t+i)$,
with $t$ denoting the iterations.
The convergence time is significantly reduced for all $t_{\mathrm{trans}}>0$, while Hamiltonian-driven
training converges after $t_{\mathrm{conv}}\sim 6600$ iterations.
The shortest convergence time for our hybrid algorithm is observed for $t_{\mathrm{trans}}=800$, which is around the minimum of the
data-driven training curve.
Numerical studies on smaller system sizes have shown similar improvements in the hybrid training approach.
While overfitting in the data-driven training starts at different points for different system sizes and
amounts of hidden neurons, \Fig{3}(c) proposes that $t_{\mathrm{trans}}$ does not need to be optimized
to ensure convergence within reasonable computational runtimes.

%
%
\section{Conclusions}
In this paper, we have explored the use of recurrent neural networks (RNNs) for studying ground state wavefunctions of
interacting Rydberg atom arrays of sizes currently accessible to experiments.
We consider RNNs both trained from projective measurement data that could be produced by a typical experiment, as well as trained 
variationally with knowledge of the target Hamiltonian.  In the latter, we show that naive variational training of RNNs becomes
challenging for arrays approaching current experimental sizes.  However, we find that RNNs which undergo a preliminary training phase
driven by a small amount of measurement data converge to accurate ground state energies with significantly less optimization steps
than RNNs optimized without access to data.
This result indicates the value of projective measurement data, obtained from Rydberg arrays and other quantum simulators,
as a mechanism to significantly improve the convergence times of variational Monte Carlo simulations.
This strategy should be immediately accessible to RNNs and other neural network wavefunction ans\"{a}tze, which are amenable to both data-driven and Hamiltonian-driven
training.

Our work also clearly indicates that the current generation of Rydberg atom arrays can produce data of high value to physicists, even at measurement rates
which may be too low to be informationally complete for a full tomographic reconstruction of the quantum state.
This strategy can be combined with the recent observation that Hamiltonian-driven optimization can also be used to mitigate errors in noisy experiments~\cite{Bennewitz2021}.
This suggests that the current generation of quantum hardware is on the cusp of bringing transformative improvement to the understanding of challenging
quantum many-body systems, by providing data to enhance variational simulation strategies for any state that can be prepared by an experimental quantum simulator.

%
\section*{Acknowledgements}
We thank J. Carrasquilla, E. Inack, I. De Vlugt, M. P. A. Fisher, D. Sels, R. Luo, and Y. H. Teoh for critically important discussions.
Simulations were made possible by the facilities of the Shared Hierarchical Academic Research Computing Network (SHARCNET) and Compute Canada. This work was supported by NSERC, the Canada Research Chair program,
the New frontiers in Research Fund, and the Perimeter Institute for Theoretical Physics. Research at Perimeter Institute is supported in part by the Government of Canada through the Department of Innovation, Science and Economic Development Canada and by the Province of Ontario through the Ministry of Economic Development, Job Creation and Trade.

\end{document}